# Investigation of Sub-configurations Reveals Stable Spin-Orbit Torque Switching Polarity in Polycrystalline Mn$_3$Sn

Boyu Zhao[1*], Zhengde Xu[1*], Xue Zhang[1], Zhenhang Kong[1], Shuyuan Shi[2†], and Zhifeng Zhu[1,3‡]

[1]School of Information Science and Technology, ShanghaiTech University, Shanghai, 201210, China

[2]Fert Beijing Institute, MIIT Key Laboratory of Spintronics, School of Integrated Circuit Science and Engineering, Beihang University, Beijing 100191, China

[3]Shanghai Engineering Research Center of Energy Efficient and Custom AI IC, Shanghai, 201210, China

## Abstract

Previous studies have demonstrated the switching of octupole moment in Mn$_3$Sn driven by spin-orbit torque (SOT). However, they have not accounted for the polycrystalline nature of the sample when explaining the switching mechanism. In this work, we use samples with various atomic orientations to capture this polycrystalline nature. We thoroughly investigate their SOT-induced spin dynamics and demonstrate that the polycrystalline structure leads to distinct outcomes. Our findings reveal that configuration II, where the Kagome plane is perpendicular to the spin polarization, exhibits robust switching with stable polarity, whereas the signals from various sub-configurations in configuration I cancel each other out. By comparing our findings with experimental results, we pinpoint the primary sources contributing to the measured AHE signals. Additionally, we establish a dynamic balance model that incorporates the unique properties of Mn$_3$Sn to elucidate these observations. Our study highlights the essential role of the polycrystalline nature in understanding SOT switching. By clarifying the underlying physical mechanisms, our work resolves the longstanding puzzle regarding the robust SOT switching observed in Mn$_3$Sn.

**Introduction**

The increasing demand for higher memory performance in modern technology has spurred extensive research into magnetic random access memory (MRAM) [1–4]. While significant advancements have been made with ferromagnetic materials [5,6], recent attention has shifted toward ferrimagnets (FiM) [7–9] or antiferromagnets (AFM) [10–17] due to their potential for faster and more energy-efficient devices. Among them, noncollinear AFM Mn$_3$Sn have emerged as particularly promising candidate [18–23]. The properties of Mn$_3$Sn are closely linked to its Kagome lattice structure [see Fig. 1(a)], which gives rise to a noncollinear spin order. As a result, despite having a negligible net magnetization, a large anomalous Hall effect (AHE) at room temperature has been demonstrated [24], which originates from the Berry curvature associated with its topological band structure. These unique properties in Mn$_3$Sn open avenues for promising spintronic applications [25,26], such as racetrack memory [27–29] and unconventional computing [30–36].

For the electrical manipulation of magnetic states in Mn$_3$Sn, it has been experimentally demonstrated that spin–orbit torque (SOT) can be used to switch the polycrystalline samples [37]. To understand the underlying switching mechanisms, the samples are classified into three different configurations based on the relative alignment between the Kagome plane, the spin polarization

($\sigma$), and the external magnetic field ($\mathbf{H}_{ext}$) [see Fig. 1(b)]. By excluding configuration III, which does not contribute to the perpendicular magnetization, it was concluded that configuration I exhibits switching behavior, whereas there is only oscillation in configuration II [37–39]. However, the switching mechanism in configuration I is still under debate [40], and subsequent works [41–43] have also identified deterministic switching in configuration II. In particular, our previous work [43] demonstrates that a strain-induced change in anisotropy is not necessarily required to achieve deterministic switching. We also show that the switching results in configuration II is insensitive to the initial state, making it more suitable for memory applications.

Although the spin dynamics of both configurations I and II have been extensively studied [43,44]. It should be aware that using only these three different configurations is insufficient to capture the polycrystalline nature of the sample [see Fig. 1(c)]. In fact, any orientation of the Kagome plane is possible, e.g., rotating the sample within the Kagome plane by 90° creates a distinct atomic environment for the spin states. To address this, we propose rotating the samples of configurations I and II through all angles in the range $\varphi$ = [0°, 360°] to capture all possible atomic states [see Fig. 2(b)]. By measuring the AHE signals, experimental studies have demonstrated robust SOT-induced switching in polycrystalline samples [7,9,17]. However, the contributions of different sub-configurations to the measured AHE signals remain unclear. In this work, by employing a more detailed methodology to examine the atomic states in polycrystalline samples, we conclude that the SOT-induced switching in configuration I will be balanced out, whereas configuration II contributes significantly to the measured AHE signals. Additionally, we propose a dynamic balance model that fully explains our results.

**Methodology**

The device studied in this work consists of Mn$_3$Sn and Pt bilayer [see Fig. 1(c)]. A charge current flowing through the heavy metal layer produces SOT [46–49] that affects the spin dynamics of Mn$_3$Sn. The Hamiltonian describing the system is expressed as:

$$\mathcal{H} = \sum_{i \neq j} A_{ij}\, \mathbf{m}_i \cdot \mathbf{m}_j + \sum_{i \neq j} \mathbf{D}_{ij} \cdot (\mathbf{m}_i \times \mathbf{m}_j) - \sum_i (\mathbf{K}_i \cdot \mathbf{m}_i)^2 - \mu_0 \mu_i \sum_i (\mathbf{m}_i \cdot \mathbf{H}_{\text{ext}}),$$

where the exchange constant $A$ = 17.53 meV, the Dzyaloshinskii-Moriya interaction (DMI) [50,51] constant $D$ = 0.833 meV, the magnetic anisotropy constant $K$ = 0.196 meV [18], and the magnetic moment $\mu = 3\mu_B$. The magnetic dynamics is obtained by solving the coupled Landau-Lifshitz-Gilbert-Slonczewski (LLGS) equations for the three Mn atoms in the same layer [52,53]:

$$\frac{d\mathbf{m}_i}{dt} = -\gamma \mathbf{m}_i \times \mathbf{H}_{\text{eff},i} + \alpha \mathbf{m}_i \times \frac{d\mathbf{m}_i}{dt} - \gamma \theta_{\text{SH}} \frac{J_c}{2eM_s d} \mathbf{m}_i \times (\mathbf{m}_i \times \sigma_i).$$

The three terms on the right-hand side represent the precession, Gilbert damping, and damping-like SOT, respectively. Here, $\gamma$ is the gyromagnetic ratio, $\alpha$ = 0.003 [37] is the Gilbert damping constant, and the spin Hall angle $\theta_{\text{SH}}$ is 0.06 [41]. The effective field $\mathbf{H}_{\text{eff},i}$ is derived from the Hamiltonian as $\mathbf{H}_{\text{eff},i} = -\frac{1}{\mu_i} \frac{\partial \mathcal{H}}{\partial \mathbf{m}_i}$, and the thickness of Mn$_3$Sn film $d$ = 30 nm [41], the saturation magnetization $M_s = 6\mu/V_{\text{cell}}$ [37] with the unit cell volume $V_{\text{cell}}$ = 0.3778 nm$^3$. The LLGS equations are solved using the fourth-order Runge-Kutta method with a time step of 5 fs.

**Results and Discussion**

As illustrated in Fig. 2(a), we rotate the Kagome plane counterclockwise by an increment angle of 1° to create 360 distinct sub-configurations denoted as $\varphi_{\text{sub\_conf}}$. Noted that the sub-configurations constructed in this procedure cover all the possible orientations that exist in the polycrystalline sample. For each sub-configuration, we then apply the same $\mathbf{J}_c$ and $\mathbf{H}_{\text{ext}}$ and observe its SOT-driven dynamics. The results are shown in Fig. 2(b), where the final state denoted by $\varphi_{\text{Final}}$ reveals six periods. Within each period, $\varphi_{\text{Final}}$ linearly increases with the rotation angle, indicating the alignment of octupole magnetic moment with the same easy axis within the Kagome plane. Towards the end of each period, a notable transition of $\varphi_{\text{Final}}$ occurs, meaning that the octupole moment is shifted to another easy axis. It is important to notice that $\varphi_{\text{Final}}$ of all the sub-configurations is restricted within −14° to 46° [marked by the blue region in the inset of Fig. 2(b)], all of which have $m_{z,\text{oct}} > 0$. Reversing the current leads to an opposite switching to $m_{z,\text{oct}} < 0$. This indicates that for configuration II, successful magnetization switching is achieved in all sample orientations, significantly contributing to the experimentally measured AHE signals. In other words, the SOT-driven switching of configuration II exhibits a stable polarity.

To further understand the switching process, we then investigate their magnetization dynamics by choosing some representative sub-configurations marked by the green and red dots. The three green dots correspond to the rotation of crystallize lattice $\varphi_{\text{sub-conf}} = 55°, 175°$, and $295°$, as sketched in the left inset of Fig. 2(c). Noted that they have an identical atomic environment, i.e., the triangle consisted of the three red atoms points in the same direction. When $\mathbf{J}_c = 5 \times 10^{10}$ A/m² and $\mathbf{H}_{\text{ext}} = 100$ Oe are applied along the +**y** axis, deterministic switching is achieved, and all the three sub-configurations are switched to the same final state with $\varphi_{\text{Final}} = -5°$ as shown in the right inset of Fig. 2(c). This reproduces our previous results [43] that the initial states with the same atomic environment will be switched to the same final state under suitable $J_c$ and $H_{ext}$. Different from the green dots that represent the early phase of a period, we then choose three sub-configurations marked by red dots that represent the late phase. As shown in the left inset of Fig. 2(d), they have the same atomic environments but different spin configuration with $\varphi_{\text{sub-conf}} = 30°$, 150°, and 270°. Under the same $J_c$ and $H_{ext}$, all of them are switched to the same final state with $\varphi_{\text{Final}} = 30°$. Comparing the switching results in Figs. 2(c) and 2(d), we can more intuitively understand the periodicity observed in Fig. 2(b). As the sample is rotated (cf. $\varphi_{\text{sub-conf}}$), the final state $\varphi_{\text{Final}}$ also rotates in the same direction, exhibiting a periodicity of 60° from −14° to 46°.

Recall that in our previous work [43], we find that the switching result is independent of the initial state. One might conclude that the results observed here can also be explained using our previous analysis. For example, under a fixed $J_c$ and $H_{ext}$, the octupole moment is switched to a fixed angle with $\varphi = 16°$. After removing $J_c$ and $H_{ext}$, the octupole moment is relaxed to the nearest easy axis. Since the easy axis rotates with the sample, it will provide a switching diagram exact the same as Fig. 2(b). However, by looking at Fig. 3(a), where $\varphi_{\text{Mid}}$ is defined as the state of octupole moment when both $J_c$ and $H_{ext}$ remain turns on, one can immediately realize that the results presented here cannot be explained using the abovementioned analysis. In contrast to the assumption that the octupole moment is fixed at $\varphi = 16°$, we find that $\varphi_{\text{Mid}}$ also rotates, and it again reveals six periods. In addition, the $\varphi_{\text{Mid}}$ covers the region marked by pink, which has different ranges from $\varphi_{\text{Final}}$ that is marked by blue. We want to emphasize that the conclusion from our previous work remains valid, i.e., for the sub-configurations with the same atomic environment, $\varphi_{\text{Final}}$ is the same regardless of the spin configuration. The switching results of the three sub-configurations in our previous work are marked by the red dots in Fig. 3(a). It is evident that our previous study represents a subset of the complete results presented here.

In our previous work, we show that stable switching of Mn₃Sn is achieved by the balance of SOT effective field $\mathbf{H}_{DL} = \boldsymbol{\sigma} \times \mathbf{m}_{oct}$ and $\mathbf{H}_{ext}$, where the handedness anomaly in the Mn₃Sn system is also included [42]. However, since $\mathbf{H}_{ext}$ is fixed while $\mathbf{H}_{DL}$ varies with $\varphi_{Mid}$, the balance of only two effective fields cannot justify the variation of $\varphi_{Final}$ and $\varphi_{Mid}$ as observed in Fig. 2(b) and Fig. 3(a). In addition, it is also important to explain why the octupole moments are restricted within a specific range. As both $\varphi_{Final}$ and $\varphi_{Mid}$ exhibit six periods, this motivates us to propose a theoretical model that includes the anisotropy energy, which also has six-fold easy axis. The anisotropy field of Mn₃Sn can be expressed as $\mathbf{H}_{an} = (\mathbf{m}_{oct} \cdot \mathbf{e})\mathbf{e} - \mathbf{m}_{oct}$, where $\mathbf{e}$ denotes the direction of the easy axis. As shown in Fig. 3(b), $\mathbf{H}_{an}$ points from the octupole moment and perpendicular to the easy axis. The $\mathbf{H}_{ext}$ and $\mathbf{H}_{DL}$ determine the approximate direction of the octupole moment, and the easy axis near this direction provides an appropriate $\mathbf{H}_{an}$. The equilibrium state of Mn₃Sn results from the vector balance among these three fields, which is sketched in Fig. 3(b). This is further supported by the relationship between $\varphi_{sub\text{-}conf}$ and $\varphi_{Diff}$, where $\varphi_{Diff}$ is defined as $|\varphi_{Final}-\varphi_{Mid}|$. As shown in Fig. 3(c), $\varphi_{Diff}$ shows a trend of being highest at both ends of the period and lowest in the middle. Since the anisotropy energy $E_{an} = -K_u(\mathbf{e} \cdot \mathbf{m}_{oct})^2$ [54], $\mathbf{m}_{oct}$ tends to align more closely with the easy axis direction to minimize $E_{an}$. As the system approaches stability, the resulting $\mathbf{H}_{an}$ will also decrease in magnitude. Therefore, we have established a dynamic theoretical model to explain the observed results. As shown in Fig. 3(d), we assume that the Mn₃Sn system is initially stabilized under the balance of $\mathbf{H}_{ext}$, $\mathbf{H}_{DL,1}$ and $\mathbf{H}_{an,1}$

(point A). The system is then rotated to point B, accompanied by the rotation of Kagome plane. It is important to note that while the direction and magnitude of $H_{ext}$ is fixed, $H_{DL}$ has a fixed magnitude but a varied direction that depends on $m_{oct}$, and $H_{an}$ is always perpendicular to the easy axis. The stability of the system now requires the balance between $H_{ext}$, $H_{DL,2}$ and $H_{an,2}$. Therefore, during the rotation of the sample from point A to point B as indicated by the black arrow, $H_{DL}$ and $H_{an}$ are adjusted accordingly to maintain the balance. The magnitude of $H_{an}$ will reach its minimum value at a specific position. Within a 60° range around this position [indicated by the orange region in Fig. 3(d)], the required $H_{an}$ to maintain the balance of the three vectors is relatively small. However, if the system is rotated out of this region, the $H_{an}$ (the dashed red arrow) that required to balance the $H_{DL}$ (the cyan dashed arrow) becomes overwhelmingly larger. This state denoted by the dashed arrows is thus unstable. Since the six easy axes of $Mn_3Sn$ are evenly distributed, one of the easy axes will always appear in the comfort region marked by orange. Once the system is rotated beyond this region, it will be realigned back to minimize the system energy. The dynamic balance model proposed here perfectly explains the variation of $\varphi_{Diff}$, which reflects $H_{an}$, and the appearance of six periods shown in Fig. 3(c). As a result, all the sub-configurations of configuration II will be switched to the easy axis within a specific region, leading to a stable switching polarity.

Similarly, the SOT induced spin dynamics in configuration I is examined by constructing the sub-configurations as shown in Fig. 4(a). Fig. 4(b) shows the relationship between $\varphi_{Final}$ and all the 360 sub-configurations under $J_c = 2.2 \times 10^{14}$ A/m² and $H_{ext} = 1000$ Oe, which is the suitable switching condition revealed in our previous work [43]. It shows that switching does not occur for the $\varphi_{\text{sub-conf}}$ within [0°, 59°], [120°, 240°] and [301°, 360°]. Fig. 4(c) shows the octupole dynamics of two $\varphi_{\text{sub-conf}}$ from these intervals, i.e., $\varphi_{\text{sub-conf}} = 160°$ and $\varphi_{\text{sub-conf}} = 340°$, where one can clearly see that the magnetic states remain unchanged. In contrast, SOT-induced switching occurs within four intervals, i.e., [60°, 90°], [91°, 119°], [241°, 269°], and [270°, 300°]. The octupole dynamics of two exemplary $\varphi_{\text{sub-conf}}$, i.e., $\varphi_{\text{sub-conf}} = 75°$ and $\varphi_{\text{sub-conf}} = 105°$, is shown in Fig. 4(d). It is intriguing to see that under the same $\mathbf{J}_c$ and $\mathbf{H}_{\text{ext}}$, $\mathbf{m}_{\text{oct}}$ is switched from +z direction to a −z for $\varphi_{\text{sub-conf}} = 75°$, whereas an opposite switching is observed for $\varphi_{\text{sub-conf}} = 105°$.

The switching results of configurations II and I are summarized into the polarity diagrams as shown in Fig. 5. In configuration II, the magnetic octupole of any sub-configuration is switched to the [−14°, 46°] interval, as indicated by the dark-colored region. Noted that all of them are directed towards the +z direction. In configuration I, the previously mentioned four switching intervals are marked by the light-colored regions in Fig. 5(b). Regions g and b with +z components are switched to regions e and d that point to −z direction, whereas regions f and c are switched from −z to +z direction. These comparable and opposite switching polarities in configuration I will be compensated, and the overall switching will not exhibit a preferred direction when these sub-configurations are evenly distributed. As a result, the spin reorientation in configuration I has a minimal contribution to the AHE signals measured in experiments.

Based on these results, now we can discuss the previous experimental findings. Ref. [17] studied SOT switching of $Mn_3Sn$ by rotating the Hall bar device, which is similar to the procedure of constructing sub-configurations conducted in our study. They have reported deterministic switching across all rotation angles. Our results show that the measured AHE signals should be primarily contributed by configuration II, whereas the signals from configuration I are weak due to their compensating characteristics.

Conclusion

By examining the different atomic orientations present in the polycrystalline $Mn_3Sn$ sample, we find that the sub-configurations in configuration II consistently contribute to a positive $m_z$ component, which is switched to the opposite state when either $\mathbf{J}_c$ or $\mathbf{H}_{ext}$ is reversed. In contrast, the signals are cancelled out in configuration I. This demonstrates that configuration II exhibits a stable polarity of SOT-driven switching, resulting in measurable AHE signals in experiments. In addition, the switching behavior reveals periodic repetition with a period of 60°. Based on this, we propose a dynamic balance model that incorporates the six-fold anisotropy present in $Mn_3Sn$. By comparing our results with experimental data, this work provides a comprehensive understanding of the switching mechanism in polycrystalline $Mn_3Sn$. Additionally, we provide guidelines for the design of efficient $Mn_3Sn$-based MRAM devices.

[*]Boyu Zhao and Zhengde Xu contributed equally to this work.

[†]Corresponding Author: smeshis@buaa.edu.cn

[‡]Corresponding Author: zzfmvp@gmail.com

The data that support the findings of this study are available from the corresponding author upon reasonable request.

**Acknowledgments**: This work was supported by National Key R&D Program of China (Grant No. 2024YFB3614100). The simulation conducted in this work is supported by HPC Platform and SIST Computing Platform at ShanghaiTech University.

**Figures**

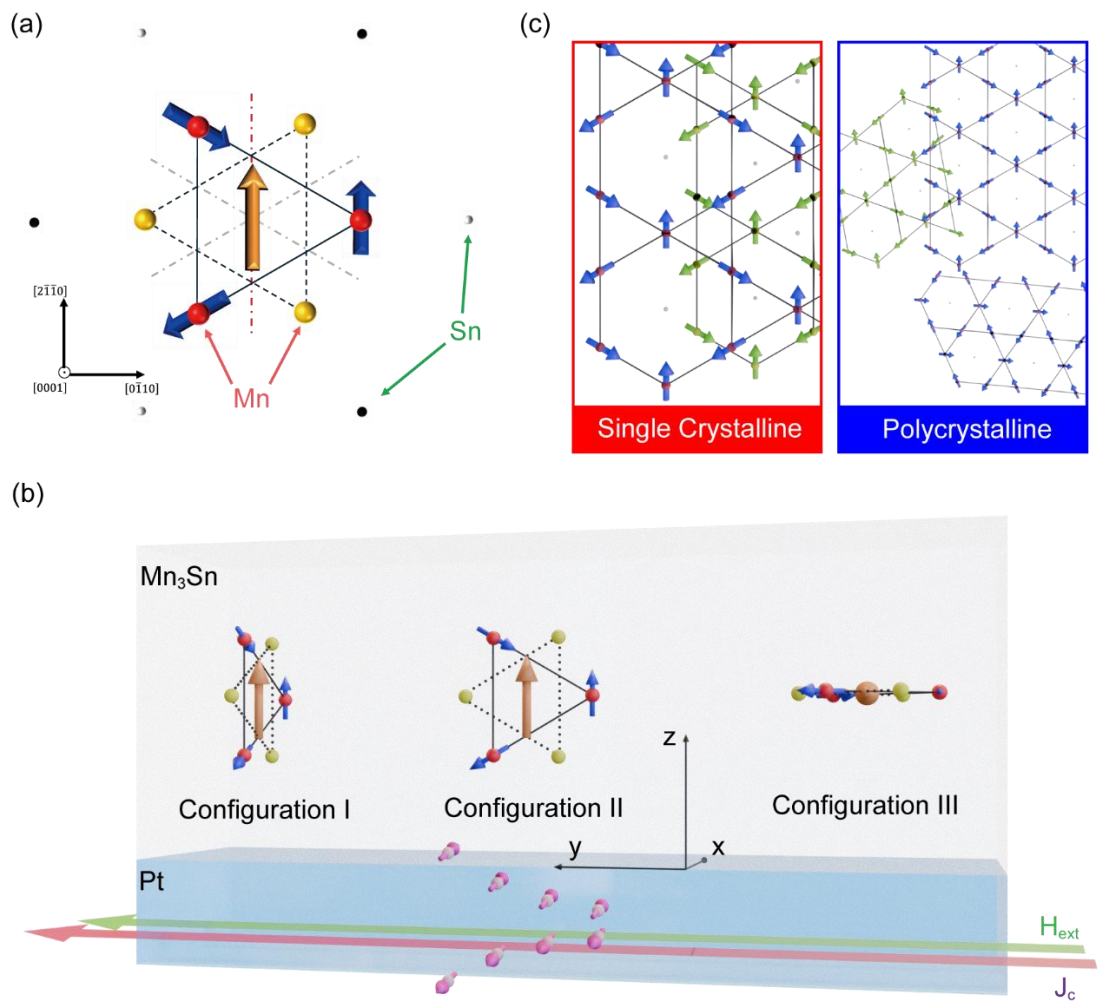

FIG. 1. (a) Atomic structure of Mn₃Sn. The red and yellow circles denote Mn atoms on the different layers. The black and silver circles denote Sn atoms on the same layer of red and yellow Mn,

respectively. Red and gray dashed lines represent easy axes of magnetic octupole. (b) The three device configurations. The Kagome plane is perpendicular to $\mathbf{H}_{ext}$ and parallel to $\sigma$ in configuration I. The Kagome plane is parallel to $\mathbf{H}_{ext}$ and perpendicular to $\sigma$ in configuration II. The Kagome plane is parallel to both $\mathbf{H}_{ext}$ and $\sigma$ in configuration III. (c) Schematic illustration of the single crystalline and polycrystalline structures.

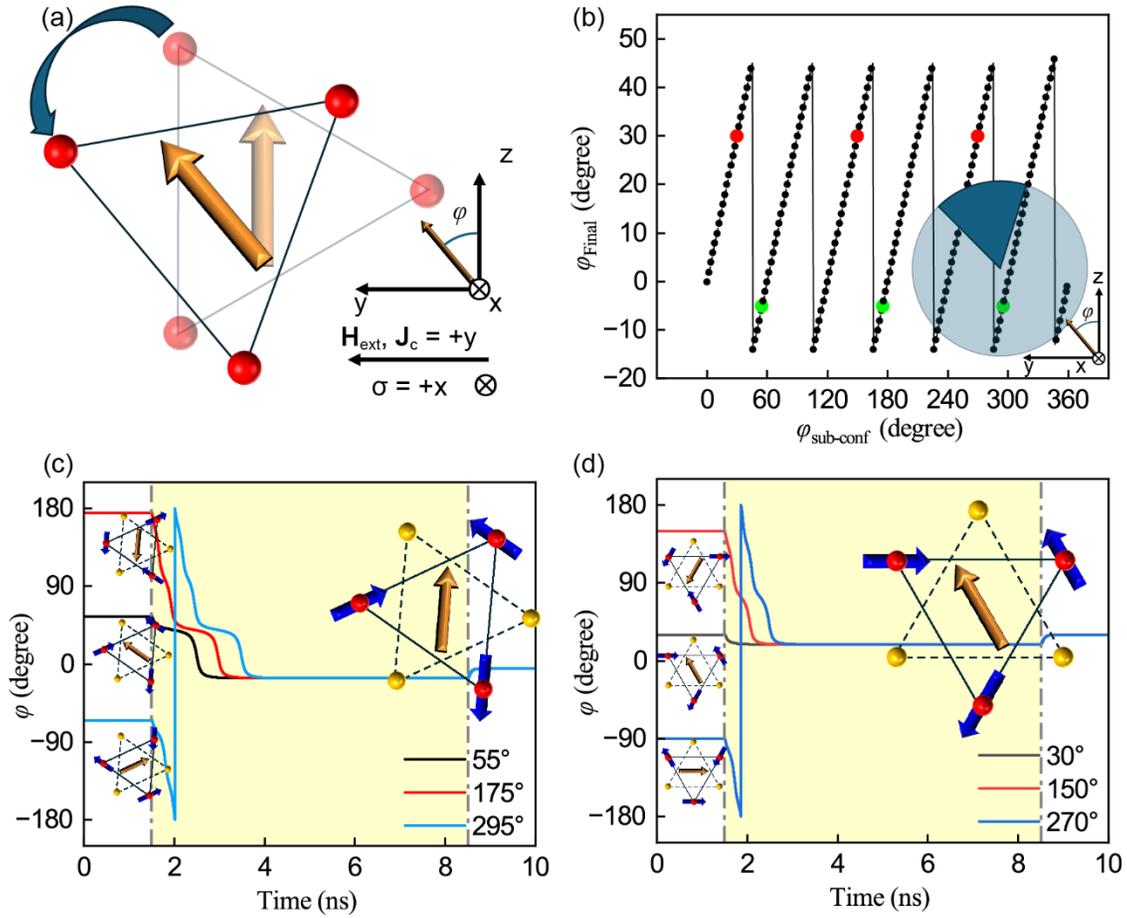

FIG. 2. (a) Starting from the +z direction with $\varphi_{sub\_conf} = 0$, sub-configurations of configuration II are constructed by rotating the sample counterclockwisely with 1° increments. At the same time, the dynamics of octupole moment is investigated under $\mathbf{J}_c = 5\times10^{10}$ A/m² and $\mathbf{H}_{ext} = 100$ Oe that are applied along the +y axis. (b) Relationship between $\varphi_{Final}$ and $\varphi_{sub\_conf}$. The dark blue region in the inset indicates the range of the final state. Dynamics of octupole moment starting from (c) $\varphi = 30°$, 150°, 270° and (d) $\varphi = 55°$, 175°, 295°, respectively.

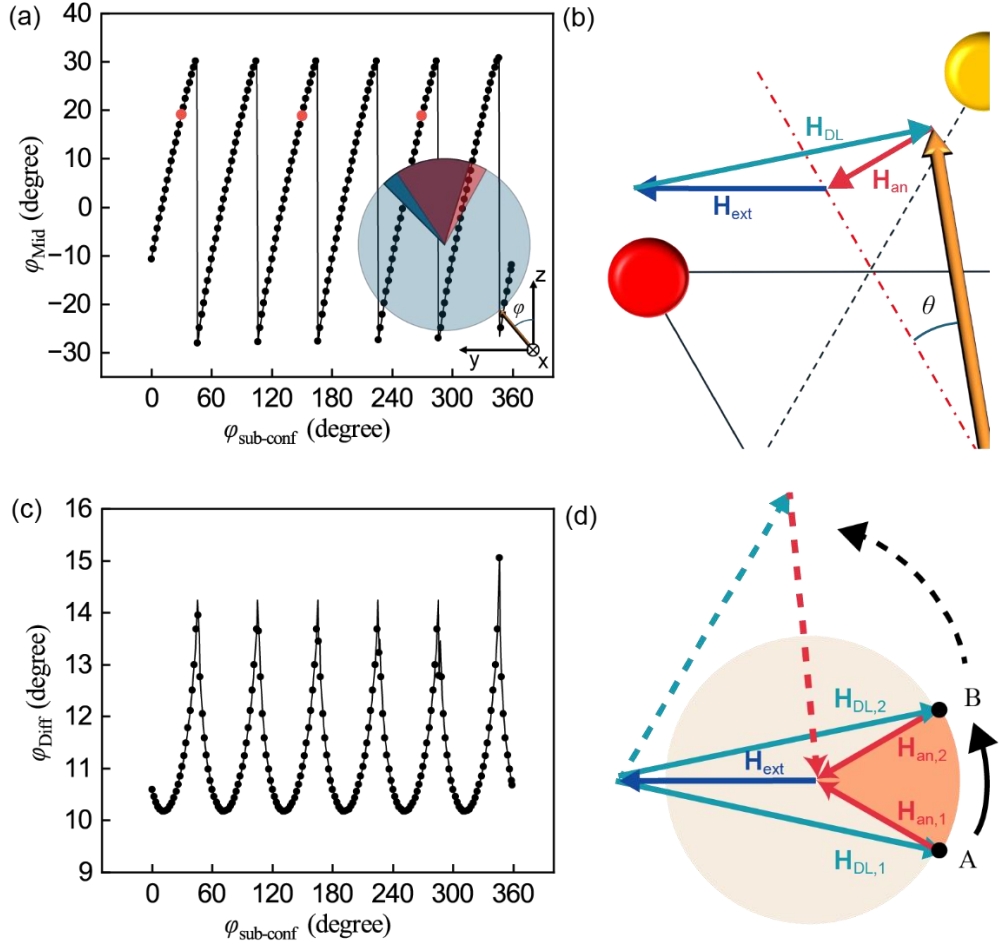

FIG. 3. (a) Relationship between $\varphi_{Mid}$ and $\varphi_{sub\_conf}$. The red region in the inset indicates the range of the middle state. (b) An equilibrium state of the magnetic octupole under $\mathbf{H}_{DL}$, $\mathbf{H}_{ext}$, and $\mathbf{H}_{an}$. The red dashed line denotes the dominant easy axis. (c) Relationship between angle difference $\varphi_{Diff}$ and $\varphi_{sub\_conf}$. (d) Dynamic balance model that integrates the three effective fields and the rotation of the sample. Note that the diagram (d) describes the effective fields, the corresponding octupole moment is denoted by the dark blue region shown in (a).

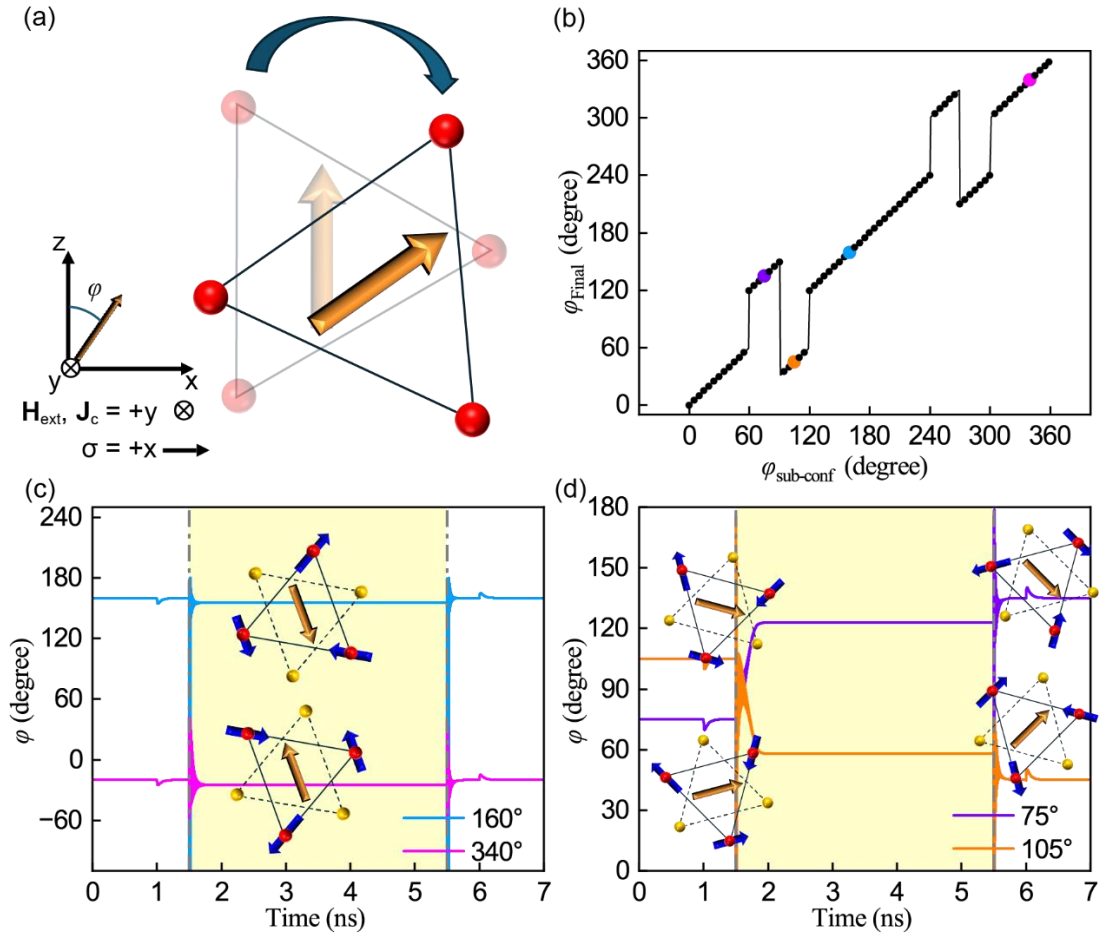

FIG. 4. (a) Construction of sub-configurations in configuration I. (b) Relationship between $\varphi_{Final}$ and $\varphi_{sub\_conf}$. Dynamics of octupole moment starting from (c) $\varphi = 160°, 340°$ and (d) $\varphi = 75°, 105°$, respectively.

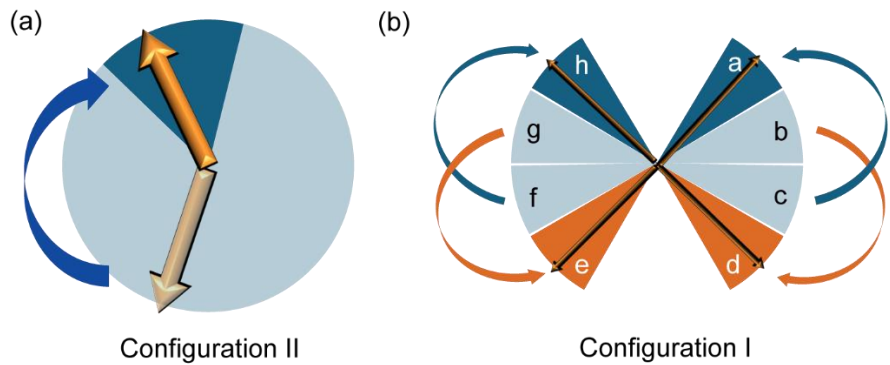

FIG. 5. Polarity diagrams of configurations (a) II and (b) I. The arrows denote the switching

direction of octupole moment.